\documentstyle[eqsecnum,preprint,aps,prd,epsf,rotate,citesort]{revtex}
\preprint{DAMTP-R-97/37,~~SU-ITP-98/11}
\date{\today}
\draft
\tighten
\begin{document}
\def\sqr#1#2{{\vcenter{\hrule height.3pt
      \hbox{\vrule width.3pt height#2pt  \kern#1pt
         \vrule width.3pt}  \hrule height.3pt}}}
\def\square{\mathchoice{\sqr67\,}{\sqr67\,}\sqr{3}{3.5}\sqr{3}{3.5}}
\def\today{\ifcase\month\or
  January\or February\or March\or April\or May\or June\or July\or
  August\or September\or October\or November\or December\fi
  \space\number\day, \number\year}

\def\Bbb{\bf}


\title{Patching up the No-Boundary Proposal \\
       with virtual Euclidean wormholes}
\author{{\sc Raphael Bousso{$^1$}}\thanks{\it bousso@stanford.edu
~~~~~~$\dagger$hac1002@damtp.cam.ac.uk} \ 
  and {\sc Andrew Chamblin{${^2}{^{\dagger}}$}}}

\address {\qquad \\ {$^1$} Dept.\ of Physics, Stanford University\\
Stanford, CA 94305-4060, USA
\qquad\\{$^2$} DAMTP, Silver Street\\
Cambridge, CB3 9EW, England
}

\maketitle

\begin{abstract}
  {\footnotesize In quantum cosmology, one often considers tunneling
    phenomena which may have occurred in the early universe.  Processes
    requiring quantum penetration of a potential barrier include black
    hole pair creation and the decay of vacuum domain walls.  Ideally,
    one calculates the rates for such processes by finding an
    instanton, or Euclidean solution of the field equations, which
    interpolates between the initial and final states.  In practice,
    however, it has become customary to calculate such amplitudes
    using the No-Boundary Proposal of Hartle and Hawking.  A criticism
    of this method is that it does not use a single path which
    interpolates between the initial and final states, but two
    disjoint instantons: One divides the probability to create the
    final state from nothing by the probability to create the initial
    state from nothing and decrees the answer to be the rate of
    tunneling from the initial to the final state.  Here, we
    demonstrate the validity of this approach by constructing
    continuous paths connecting the ingoing and outgoing data, which
    may be viewed as perturbations of the set of disconnected
    instantons.  They are off-shell, but will still dominate the path
    integral as they have action arbitrarily close to the no-boundary
    action.  In this picture, a virtual domain wall, or wormhole, is
    created and annihilated in such a way as to interface between the
    disjoint instantons.  Decay rates calculated using our
    construction differ from decay rates calculated using the
    No-Boundary Proposal only in the prefactor; the exponent, which
    usually dominates the result, remains unchanged.}
\end{abstract}
\pagebreak

\section{Introduction}

\begin{center}
\small {Ripple in still water}\\
\small {When there is no pebble tossed}\\
\small {Nor wind to blow}\\
\small {\it -The Grateful Dead}\\
\end{center}

Originally \cite{HarHaw83}, the No-Boundary Proposal was an
attempt to eliminate the initial and final singularities (which are
`boundaries' for spacetime in cosmological scenarios), by
considering the universe as a history in imaginary time.  One is then
led to a picture of a universe which is finite in imaginary time and
`without boundary'.  This picture involves both `real time'
(or Lorentzian) and `imaginary time' (or Euclidean) sections of the
full complexified spacetime.  When one combines this picture with the
sum-over-histories prescription for calculating amplitudes, it is
natural to think of a Euclidean section (which matches smoothly to a
Lorentzian section across a spacelike three-surface of vanishing
extrinsic curvature) as an `instanton', which mediates the creation of
the Lorentzian section from `nothing'.  In this way, 
a new universe can appear from `nothingness', even though there was no
source around to precipitate such an event.  One then speaks of
creating universes from nothing.  Similarly, one can consider the time
reverse and speak of universes `annihilating' to
nothingness.  One can calculate Euclidean actions of the relevant
instantons and obtain the rate at which various universes appear
from nothing.  One can thus `predict' the most likely initial
state for the universe.  The key point about this version of
the No-Boundary Proposal is that it tells you how to calculate the
probability that {\it something} appears from {\it nothing}.

Here, we are concerned with another variant of the
No-Boundary Proposal, which tells you how to calculate the rate at
which {\it something} decays into {\it something else}.  This version
of the proposal is often applied to the study of black hole pair
creation.  Throughout this paper we use units 
in which $\hbar = c = G = 1$.

\section{Approaches to Gravitational Tunneling}

The pair creation of black holes, first discovered by Gibbons
\cite{gazzorigin}, has been studied enthusiastically for
a number of years. It corresponds to a non-perturbative, topological
fluctuation of the gravitational field. As such it is one of the few
effects of quantum gravity that one can hope to study quantitatively
in a semi-classical approximation. It has been used to
investigate the entropy of black
holes~\cite{GarGid94,DowGau94b,HawHor95}, and electric-magnetic
duality in quantum gravity~\cite{HawRos95b}.  In the cosmological
context, it has clarified
the important role of the Hartle-Hawking No-Boundary Proposal in
quantum gravity~\cite{BouHaw95,BouHaw96}, and it may have profound
consequences for the global structure of the universe~\cite{Bou98}.

\subsection{Bounce Approach} \label{ssec-instapproach}

Black hole pair creation can be analyzed semi-classically by the use
of instanton methods. Typically, the nucleation process is described
by a single Euclidean solution of the Einstein equations, a bounce. It
interpolates between an initial spacelike section without black holes,
and a final spacelike section containing black holes, bouncing back to
the initial section in a time-symmetric fashion. An instanton is half
a bounce, i.e.\ a geometry connecting initial and final spacelike
sections, but not bouncing back. One calculates the bounce action,
$I_{\rm pc}$, which must be renormalized by subtracting off the action
$I_{\rm bg}$ of a Euclidean geometry containing only background
spacelike sections. Thus one obtains the pair creation rate $\Gamma$:
\begin{equation}
\Gamma =
\exp \left[ - \left( I_{\rm pc} - I_{\rm bg} \right) \right],
\label{eq-pcr-usual}
\end{equation}
where we neglect a prefactor. Note that both $I_{\rm pc}$ and $I_{\rm
  bg}$ are typically infinite, but their difference can be made well
defined and finite in a suitable limit.  This prescription has been
used very successfully by a number of authors
\cite{GarGid94,DowGau94a,DowGau94b,HawHor95,HawRos95b,Ros94,Ros95} for
the pair creation of black holes on various backgrounds. It is
motivated by analogies in quantum mechanics and quantum field
theory~\cite{Col77,CalCol77}, as well as by considerations of black hole
entropy~\cite{GarGid94,DowGau94b,HawHor95}.

\subsection{Quantum Cosmological Approach} \label{ssec-qcapproach}

A positive cosmological constant, or a domain wall,
typically causes the universe to close. In these situations, there are
Lorentzian solutions with and without black holes, but there are no
known Euclidean solutions connecting their spacelike slices.
Instead, there are two separate, compact Euclidean geometries
corresponding to creation from nothing of a universe with, and
without a black hole pair.  Thus the instanton technique outlined
above could not be used directly. Before describing how
virtual domain walls mend this problem, we review how the
problem was circumvented using concepts from quantum cosmology.

In quantum cosmology one works with the concept of the {\em wave
  function of the universe}. The wave function takes different values
for a universe with, and without black holes. 
The squared amplitude of the wave function yields a probability
measure.  According to the Hartle-Hawking No-Boundary Proposal, the
wave function of the universe, evaluated for a specified
three-geometry, is given by a path integral over all closed,
smooth complex geometries that match the specified boundary conditions
on the spacelike section, and have no other boundary; the integrand is
the exponential of minus the Euclidean action of the geometry. In
the semi-classical approximation, the wave function is
approximated as
\begin{equation}
\Psi = e^{-I_{\rm inst}},
\end{equation}
where $I_{\rm inst}$ is the action of a saddlepoint solution which
satisfies the Einstein equations under the given boundary conditions.
If there is no such instanton, the wave function will be zero
semi-classically; if there are several, they have to be summed over.
The probability measure for a given universe is thus related to the
action of an instanton which describes the nucleation of the universe
from nothing:

\begin{equation}
P = \Psi^* \Psi = e^{-2 I^{\rm Re}_{\rm inst}}.
\label{eq-P}
\end{equation}
Clearly, the probability depends only on the real part of the
instanton action, $I^{\rm Re}_{\rm inst}$.
We shall see in the next section that there are, indeed, two
instantons, each of which nucleates a universe from
nothing. One will lead to spacelike sections with black holes, the
other to an empty background universe. Because of the cosmological
term, the Euclidean geometry is compact, and the actions of both
instantons are finite.  Thus probability measures can be assigned to a
universe with, and without black holes.

But how are these probability measures related to pair creation? It
seems that all one can do in quantum cosmology is to compare the
probability measure for a universe with {\em one} pair of black holes
to that of an empty universe. The black hole instanton would then be
without any cosmological relevance whatsoever -- it could only produce
a single black hole pair in an exponentially large universe. It is
possible, however, to propose the following approach \cite{BouHaw96}:
Consider an arbitrary Hubble volume in an inflating universe.
Typically, this volume will not contain black holes; it will be
similar to a Hubble volume of de~Sitter space. After one Hubble time,
its spatial volume will have increased by a factor of $e^3 \approx
20$. By the de~Sitter no hair theorem, one can regard each of these
$20$ Hubble volumes as having been nucleated
independently~\cite{GarLin94}, through either the empty, or the black
hole instanton. Thus, one allows for black hole
pair creation, since some of the new Hubble volumes may contain black
holes.  We shall see later that the spacelike sections can be taken to
be three-spheres in the case of a universe without black holes, and $
S^1 \times S^2 $ for a universe with a single pair of black holes. We
will therefore compare the wave functions for spacelike slices with
these two topologies.  Using the No-Boundary
Proposal~\cite{HarHaw83}, one can assign probability measures to both
instanton types. The ratio of the probability measures,
\begin{equation}
\Gamma = \frac{P_{\rm BH}}{P_{\rm no\, BH}},
\label{eq-pcr-qc}
\end{equation}
reflects the ratio of the number of Hubble volumes
containing black holes, to the number of empty Hubble
volumes. This is true as long as this ratio is very
small, so that the holes will be widely separated. But we shall
see that this condition is satisfied whenever the semi-classical
approximation is valid. Since this argument applies to every new
generation of Hubble volumes, the ratio $\Gamma$ is the number of
black hole pairs produced per Hubble volume, per Hubble time. In other
words, $\Gamma$ is the rate of black hole pair creation in inflation.

\subsection{The Cosmological Pair Creation Instantons}

\begin{figure}[htb]
\epsfxsize=\textwidth
\epsfbox{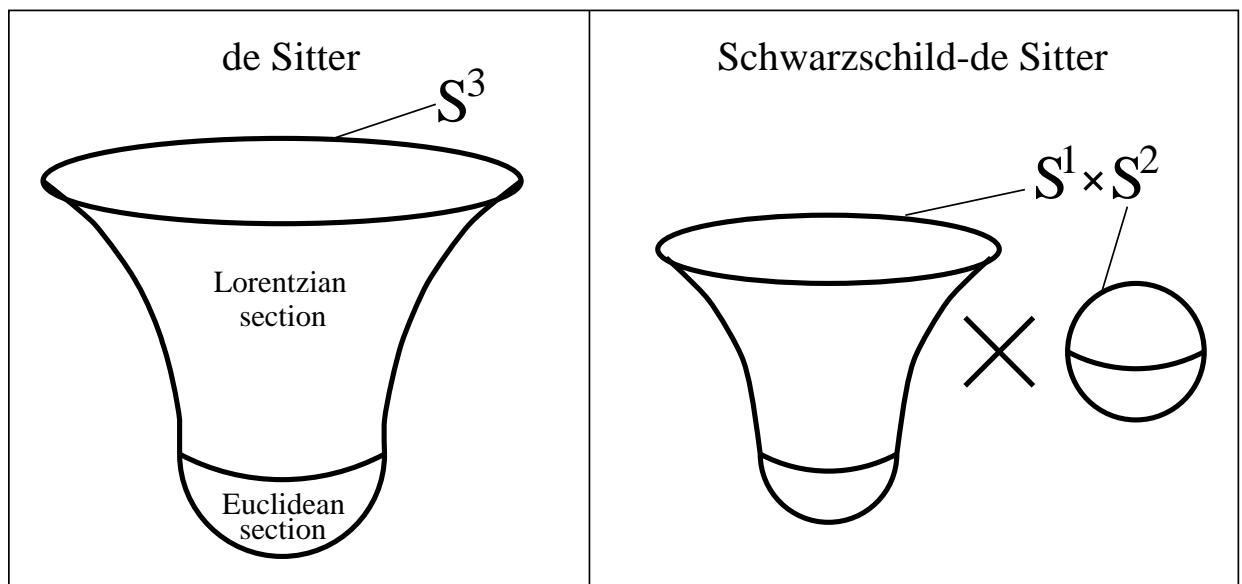}
\caption[]%
{\small\sl The creation of a de Sitter universe (left) can
  be visualized as half of a Euclidean four-sphere joined to a
  Lorentzian four-hyperboloid. The picture on the right shows the
  corresponding nucleation process for a de Sitter universe containing
  a pair of black holes. In this case the spacelike slices have
  non-trivial topology.}
\label{fig-tun}
\end{figure}

We now illustrate this approach by reviewing its implementation
for the case of neutral black holes created on a cosmological
background~\cite{GinPer83,BouHaw95,BouHaw96}.  We begin with the
simpler of the two spacetimes, an inflationary universe without black
holes, where the spacelike sections are round three-spheres. In
the Euclidean de Sitter solution, the three-spheres begin at zero
radius, expand and then contract in Euclidean time. Thus they form a
four-sphere of radius $\sqrt{3/\Lambda}$.  The analytic continuation
can be visualized (see Fig.~\ref{fig-tun}) as cutting the four-sphere
in half, and then joining to it half the Lorentzian de~Sitter
hyperboloid.  The real part of the Euclidean action for this
geometry comes from the half-four-sphere only: $I^{\rm
  Re}_{\rm no\, BH} = - 3\pi/2\Lambda$.  Thus, the
probability measure for de~Sitter space is
\begin{equation}
P_{\rm no\, BH} = \exp \left( \frac{3\pi}{\Lambda} \right).
\end{equation}

\begin{figure}[htb]
\epsfxsize=\textwidth
\epsfbox{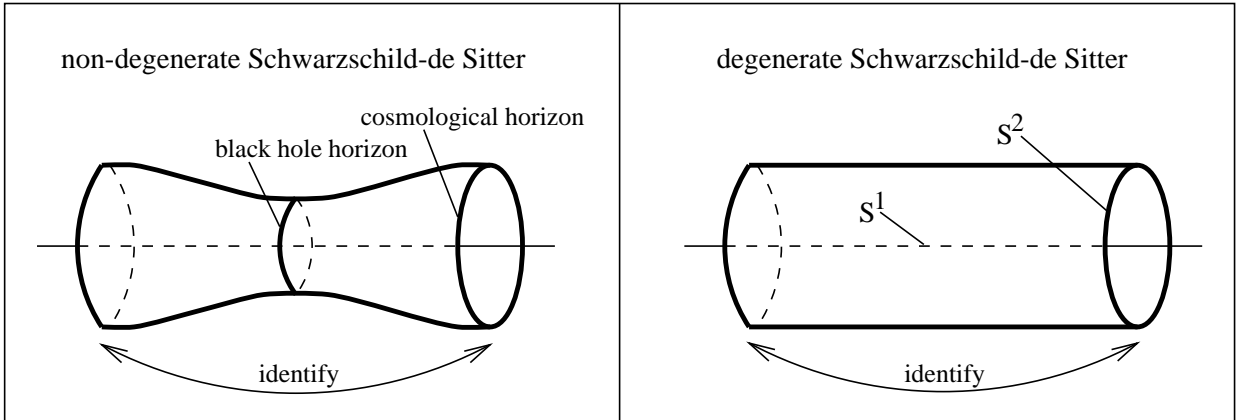}
\caption[Spacelike sections of Schwarzschild-de~Sitter space]%
{\small\sl The spacelike slices of Schwarzschild-de~Sitter space have
  the topology $S^1 \times S^2$. In general (left), the size of the
  two-sphere varies along the one-sphere. If the black hole mass is
  maximal, however, all the two-spheres have the same size (right).
  Only in this case is a smooth Euclidean solution admitted.}
\label{fig-space-secs}
\end{figure}

Now we need to go through the same procedure with the
Schwarzschild-de~Sitter solution, which corresponds to a pair of black
holes immersed in de~Sitter space. Its Lorentzian metric is given by
\begin{equation}
ds^2 = -V(r) dt^2 + V(r)^{-1} dr^2 + r^2 d\Omega^2,
\end{equation}
where
\begin{equation}
V(r) = 1 - \frac{2\mu}{r} - \frac{\Lambda}{3} r^2.
\end{equation}
Here $\mu$ parameterizes the mass of the black hole, and for $\mu=0$
the metric reduces to de~Sitter space.  The spacelike sections have
the topology $S^1 \times S^2$.  This can be seen by the following
analogy: Empty Minkowski space has spacelike sections of topology
${\rm {\bf R}}^3$. Inserting a black hole changes the topology to $S^2
\times {\rm {\bf R}}$.  Similarly, if we start with de~Sitter space
(topology $S^3$), inserting a black hole is like punching a hole
through the three-sphere, thus changing the topology to $S^1 \times
S^2$.  In general, the radius of the $S^2$ varies along the $S^1$. In
the static slicing of Schwarzschild-de~Sitter, the maximum two-sphere
corresponds to the cosmological horizon, the minimum to the black hole
horizon.  This is shown in Fig.~\ref{fig-space-secs}.

What we need is a Euclidean solution that can be analytically
continued to contain this kind of spacelike slice.  It turns out that
such a smooth instanton does not exist in general for the Lorentzian
Schwarzschild-de~Sitter spacetimes. The only exception is the
degenerate case, where the black hole has the maximum possible size,
and the radius of the two-spheres is constant along the $S^1$ (see
Fig.~\ref{fig-space-secs}). The corresponding Euclidean solution is
just the topological product of two round two-spheres, both of radius
$1/\sqrt{\Lambda}$~\cite{GinPer83}.  It can be analytically continued
to the Lorentzian Schwarzschild-de~Sitter solution by cutting one of
the two-spheres in half, and joining to it the 2-dimensional
hyperboloid of $1+1$ dimensional Lorentzian de~Sitter space, as shown
in Fig.~\ref{fig-tun}.  In the Lorentzian regime the $S^1$ expands
exponentially, while the two-sphere just retains its constant radius.
Thus, unless more sophisticated boundary conditions are
employed~\cite{BouHaw98}, the Euclidean approach predicts the
nucleation of black holes of the maximum size, $r_{\rm BH} =
\Lambda^{-1/2}$.

The real part of the Euclidean action for this instanton is given by
$I^{\rm Re}_{\rm BH} = - \pi/\Lambda$, and the corresponding
probability measure is
\begin{equation}
P_{\rm BH} = \exp \left( \frac{2\pi}{\Lambda} \right).
\end{equation}
Now we can take the ratio of the two probability measures, and obtain
the pair creation rate: 
\begin{equation}
\Gamma = \exp \left( -\frac{\pi}{\Lambda} \right).
\end{equation}

This example illustrates the analogy between the standard prescription
for pair creation, Eq.~\ref{eq-pcr-usual}, and the result obtained from
the No-Boundary Proposal: By Eqs.~(\ref{eq-pcr-qc})
and~(\ref{eq-P}),
\begin{equation} 
\Gamma = 
\frac{P_{\rm BH}}{P_{\rm no\, BH}} =
\exp \left[ - \left( 2I^{\rm Re}_{\rm BH} - 2I^{\rm Re}_{\rm no\, BH}
  \right) \right],
\label{eq-pcr-unified}
\end{equation}
where $I^{\rm Re}$ denotes the real part of the Euclidean action of a
nucleation geometry. But we have seen that the only contribution to
$I^{\rm Re}$ comes from the action of the Euclidean sector of the
nucleation geometry, the instanton. This, in turn, is equal to half of
the action of the complete bounce solution, which is used in the usual
pair creation framework. Thus $I_{\rm pc} = 2I^{\rm Re}_{S^1 \times
  S^2}$ and $I_{\rm bg} = 2I^{\rm Re}_{S^3}$, and we recover
Eq.~\ref{eq-pcr-usual}.

The quantum cosmological approach to black hole pair creation outlined
above clearly differs from the usual bounce approach. In the latter, a
Euclidean time region smoothly interpolates between the two different
spacelike slices; in quantum cosmology, on the other hand, one can
think of the pair creation process as the annihilation of a Hubble
volume, and its subsequent recreation with a different spatial
topology.  Thus it is, perhaps, less obvious to see the analogy to
quantum mechanical tunneling instantons that motivate the bounce
approach to pair creation. Some have therefore regarded cosmological
and domain wall pair creation scenarios with a high degree of
scepticism.  While we have always found the probabilistic argument
given above convincing, and thought it justified to calculate a pair
creation rate from the No-Boundary Proposal, we hope to allay any
further worries by explicitly constructing an interpolating tunneling
path using virtual wormholes, or domain walls. It will be shown that
the wormhole contribution to the Euclidean action can be negligible.
The quantum cosmological formula for the pair creation rate,
Eq.~(\ref{eq-pcr-qc}), will thus be confirmed.

\subsection{The Patching Proposal}

As we have discussed above, when one uses the No-Boundary Proposal to
calculate a tunneling amplitude one does not actually construct an
imaginary time resonance connecting the initial and final states.  In
fact, in many scenarios involving topology change, there simply {\it
  does not exist} any globally regular solution of the relevant
Euclidean equations of motion which interpolates between ingoing and
outgoing data.  There are thus two problems: First, if one takes the
point of view that disconnected geometries should not be allowed in
the path integral, one would have to exclude the solutions given in
the previous section. Then it seems that there would be no saddle
point, and the transition rate should vanish semi-classically. Second,
even if disconnected geometries are admitted, it is not immediately
obvious why the actions of the two disjoint instantons should be
subtracted, rather than added. Here we make a proposal that solves
both problems: the idea is to join the disconnected geometries by
small virtual domain walls.

In the path integral approach, one obtains an amplitude by summing
over {\it all} paths, whether they are solutions or not. Given a path
integral with one saddlepoint, it will be of interest to consider the
consequences of taking the integral over all paths {\em except} for
the saddlepoint and very small perturbations about it. If the removed
region is sufficiently small, there will still be a region of
stationary action, located around the excised area. There the
oscillations of the integrand will not be destructive and the
amplitudes will add up. Paths which are close to being solutions will
then dominate the sum.

To make this precise, let us assume that the saddlepoint action is
given by $I_0$, and consider perturbations of this solution
parameterized by $\delta$. Then the action near the saddlepoint will
be given by
\begin{equation}
I(\delta) = I_0 + \frac{1}{2} \rho \delta^2,
\end{equation}
where $\rho$ is the second derivative of the action evaluated at the
saddlepoint. Ignoring other perturbations, the path integral will be
given by
\begin{eqnarray}
\int_{-\infty}^{\infty} d\delta\, e^{-I}
 & = &  e^{-I_0} \int_{-\infty}^{\infty} d\delta\,  e^{-\frac{1}{2} \rho
  \delta^2 } 
\label{eq-pathint} \\
& = & \sqrt{\frac{2\pi}{\rho}} e^{-I_0}
\label{eq-prefexp}
\end{eqnarray}
in the saddlepoint approximation.

Our idea is the following: We take the saddlepoint geometry to be the
combination of a half-four-sphere (which annihilates a de~Sitter
Hubble volume) with half of an $S^2 \times S^2$, which will create
Schwarzschild-de~Sitter space from nothing. One may connect these two
disjoint instantons by removing a small four-ball of radius $\delta$
on each and joining them together on the resulting boundaries. This
leads to a family of near-solutions, in which the instantons are
connected though a virtual domain wall of size $\delta$. These
geometries, which violate Einstein's equations (with positive
energy sources) in a small region, will
actually interpolate between the initial and final spacelike sections.
The disjoint saddlepoint solution is recovered in the limit where
$\delta \rightarrow 0$. The idea works just the same for the disjoint
instantons associated with black hole pair creation on (real) domain
walls, which will be discussed below.

One must assume that the virtual domain walls may not be smaller than
Planck size: $|\delta|>1$. Forbidding the disconnected geometries thus
corresponds to removing the region $|\delta|<1$ from the range of
integration in Eq.~(\ref{eq-pathint}). By calculating the action
contribution of the virtual domain wall, we will show that $\rho$ is
of order one. Thus we are excising a region of about one standard
deviation from the integral. This will reduce the prefactor of the
wave function to about a third of its value in Eq.~(\ref{eq-prefexp}).
The exponent, which typically is much more significant, will not
change at all.  
(Since $\delta$ should be small compared to the size of the instanton,
there will strictly also be an upper bound.  But the overwhelming
contribution to the integral comes from the first few standard deviations.
Therefore the error from integrating to infinity will be negligible except
for Planck scale instantons, when the semiclassical approach breaks down
anyway).

Therefore, connected geometries will dominate the path integral in the
absence of disconnected ones. This solves the first problem raised at
the beginning of this subsection. The second problem is resolved by
the change of orientation at the virtual domain wall, which will cause
the two instanton actions to enter the exponent with opposite sign.

The construction of the off-shell interpolating Euclidean paths will
be presented in detail in Sec.~\ref{sec-interpol}.  In order to be
precise, we will explicitly go through our construction for the
scenario where black holes are pair produced in the background of a
Vilenkin-Ipser-Sikivie domain wall, as discussed in \cite{cald}.  Our
motivation for treating the tunneling process of black hole pair
production in the presence of a domain wall is twofold.  First of all,
we are going to have to introduce the notion of a domain wall, or
infinitely thin wormhole, anyway in order to implement our proposal.
Second of all, the qualitative features of black hole pair production
by a domain wall are identical to those of black hole pair production
in a de Sitter background; indeed, it will not be hard to see that
everything we will say here for the domain wall situation will go
through for the de Sitter situation, where one is interested in the
creation of primordial black holes in the early universe
\cite{BouHaw95,BouHaw96,BouHaw97b}.  With all of this in mind, we now
present a remedial overview of domain walls.

\section{Black Hole Pair Creation on Domain Walls}

\subsection{Domain Walls: A Brief Introduction}

A vacuum domain wall is a two-dimensional topological defect which can
form whenever there is a breaking of a discrete symmetry.  Commonly,
one thinks of the symmetry breaking in terms of some Higgs field
$\Phi$.  If ${\cal M}_{0}$ denotes the `vacuum manifold' of $\Phi$
(i.e., the submanifold of the Higgs field configuration space on which
the Higgs acquires a vacuum expectation value because it will minimize
the potential energy $V(\Phi)$), then a necessary condition for a
domain wall to exist is that ${\pi}_{0}({\cal M}_{0}) \not= 0$.  In
other words, vacuum domain walls arise whenever the vacuum manifold is
not connected.  The simplest example of a potential energy which gives
rise to vacuum domain walls is the classic `double well' potential,
which is discussed in detail (along with many related things) in
\cite{pod}.

(Note: In general, domain walls can arise as (D-2)-dimensional defects
(or extended objects) in D-dimensional spacetimes.  In fact, domain
walls are a common feature in the menagerie of objects which can arise
in the low-energy limit of string theory, as has been discussed in
detail in \cite{paul} and \cite{cvet}.)

>From what we have said so far, the Lagrangian density for the matter
field $\Phi$ is given as \cite{pod}
\begin{equation}
{\cal L}_{m} = 
-{\frac{1}{2}}g^{{\alpha}{\beta}}{\partial}_{\alpha}
{\Phi}{\partial}_{\beta}{\Phi} -
V(\Phi).
\end{equation}

The exact form of $V(\Phi)$ is not terribly important.  All that we
require in order for domain walls to be present is that $V(\Phi)$ has
a discrete set of degenerate minima, where the potential vanishes.
Given this matter content, the full (Lorentzian) Einstein-matter
action then reads:
\begin{equation}
S = {\int}_{\!\!M} d^4\! x \, \sqrt{-g}\,
\Big[ {R \over 16 \pi} + {\cal L}_{m} \Big] + 
\frac{1}{8{\pi}}{\int}_{\!\!\partial M} d^{3}\!x\, \sqrt{h}K.
\end{equation}

Here, $M$ denotes the four-volume of the system, and $\partial M$
denotes the boundary of this region.  One obtains the Euclidean
action, $I$, for the Euclidean section of this configuration by
analytically continuing the metric and fields and reversing the
overall sign.  The `simplified' form of this Euclidean action in the
thin wall limit has been derived in a number of recent papers
(\cite{cald}, \cite{shawn1}) and so we will not reproduce the full
argument here.  Basically, one first {\it assumes} that there is no
cosmological constant ($R = 0$) and then one uses the fact that the
fields appearing in the matter field Lagrangian depend only on the
coordinate `$z$' normal to the wall, and one integrates out this
$z$-dependence to obtain the expression
\begin{equation}
I = - \sum_{i=1}^{n}
 {\frac{\sigma_{i}}{2}}{\int}_{\!\!D_{i}}d^{3}\!x \sqrt{h_{i}}.
\end{equation}

Here, $D_i$ denotes the $i$-th domain wall, ${\sigma}_i$ is the energy
density of the domain wall $D_i$, $h_i$ is the determinant of the
three-dimensional metric ${h^{ab}}_{(i)}$ induced on the domain wall
$D_i$ and $n$ is the total number of domain walls.  Now, it is well
known that variation relative to ${h^{ab}}_{(i)}$ on each domain wall
will yield the Israel matching conditions.  Since we will make use of
these conditions, we reproduce them here for the convenience of the
reader:
\begin{enumerate}
\item A domain wall hypersurface is totally umbilic, i.e., the second
  fundamental form $K_{ij}$ is proportional to the induced metric
  $h_{ij}$ on each domain wall world sheet.
\item The discontinuity in the second fundamental form on each domain
  wall hypersurface is $[K_{ij}]_\pm = 4 \pi \sigma h_{ij}$.
\end{enumerate}

Thus, the energy density of a thin domain wall measures the jump in
the extrinsic curvature of surfaces parallel to the wall as one moves
through the wall.  We will use these conditions to do quick
`cut-and-paste' constructions of virtual domain wall surfaces.

Now, the above the discussion is a nice summary of the field
theoretical aspects of a generic vacuum domain wall, but what would a
gravitating domain wall actually look like?

\subsection{The VIS Domain Wall Spacetime}

Solutions for the gravitational field of a domain wall were found by
Vilenkin \cite{vil} (for an open wall) and Ipser and Sikivie \cite{ip}
(for closed walls).  The global structure of these
Vilenkin-Ipser-Sikivie (or `VIS') domain walls has been extensively
discussed recently (\cite{cald}, \cite{shawn1}) so we will only
present a brief sketch here.

To start with, we are looking for a solution of the Einstein equations
where the source term is an energy momentum tensor describing a
distributional source located at $z = 0$:
\begin{equation}
T_{\mu \nu} = \sigma\, \delta(x)\, {\rm diag} (1,1,1,0).
\label{stressenergy}
\end{equation}

It is not possible to find a static solution of the Einstein equations
with this source term; indeed, the VIS solution is a time-dependent
solution describing a uniformly accelerating domain wall.  In order to
understand the global causal structure of the VIS domain wall, it is
most useful to use coordinates $(t, x, y, z)$ so that the metric takes
the form
\begin{equation} 
ds^{2} = \Big(1 - k|z|\Big)^2
dt^{2} - dz^{2} - \Big(1 - k|z|\Big)^2 
e^{2kt}   (dy^{2} + dx^{2}).
\label{vismetric}
\end{equation} 

Here, $k = 2{\pi}{\sigma}$.  The gravitational field of this solution
has unexpected properties.  For example, in the Newtonian limit of the
Einstein equations for (\ref{vismetric}) one obtains the equation
\[
{\nabla}^{2}{\phi} = -2{\pi}{\sigma},
\]
\noindent where $\phi$ is the Newtonian gravitational 
potential and $\sigma$ is the energy density of the wall.  From this
equation it is clear that a wall with {\it positive} surface energy
density will have a repulsive gravitational field, whereas a wall with
negative energy density will have an attractive gravitational field.
An even simpler way to see that the (positive $\sigma$) VIS wall is
repulsive is to notice that the $t-z$ part of the metric is just the
Rindler metric.

Further information is recovered by noticing that the $z$=constant
hypersurfaces are all {\it isometric} to $2+1$ dimensional de Sitter
space:
\begin{equation} 
ds^{2} = dt^{2} - e^{2kt}(dy^{2} + dx^{2}). 
\end{equation} 

Given that $2+1$ de Sitter has the topology ${\rm S}^{2} \times {\Bbb
  R}$ it follows that the domain wall world sheet has this topology.
In other words, at each instant of time the domain wall is
topologically a two-dimensional sphere.  Indeed, in the original
Ipser-Sikivie paper a coordinate transformation was found which takes
the $(t, x, y, z)$ coordinates to new coordinates $(T, X, Y, Z)$ such
that in the new coordinates the metric becomes (on each side of the
domain wall):
\begin{equation} 
ds^{2} = dT^{2} - dX^{2} - dY^{2} - dZ^{2}.  
\end{equation}

\noindent Furthermore the domain wall, 
which in the old coordinates is a plane located at $z = 0$, is in the
new coordinates the hyperboloid
\begin{equation} 
X^{2} + Y^{2} + Z^{2} = {1\over {k}^2} + T^{2}.  
\label{dwhyper} 
\end{equation} 

\begin{figure}[htb]
\hspace*{\fill} \vbox{\epsfxsize=11cm
\rotate[r]{\epsfbox{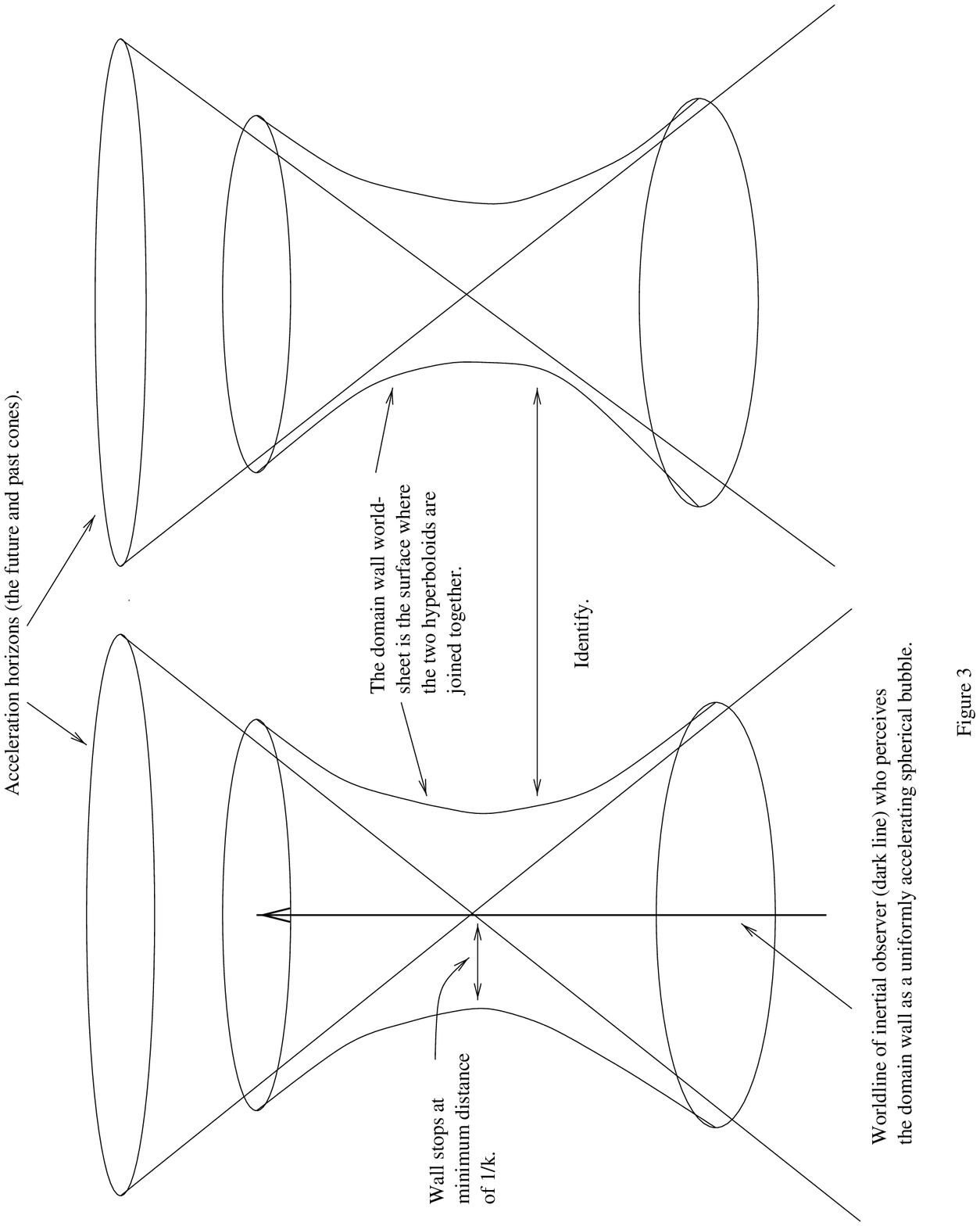}}} \hspace*{\fill}
\caption[Causal structure of VIS domain wall spacetime.]
{\small\sl}
\label{vis}
\end{figure}

Of course, the metric induced on a hyperboloid embedded in Minkowski
spacetime is just the de Sitter metric, and so this is consistent with
what we have already noted.  This metric provides us with a useful way
of constructing the maximal extension of the domain wall spacetime:

\noindent First, take two
copies of Minkowski space, and in each copy consider the interior of
the hyperboloid determined by equation (\ref{dwhyper}). Then match
these solid hyperboloids to each other across their respective
boundaries; there will be a ridge of curvature (much like the edge of
of a lens) along the matching surface, where the domain wall is
located.  Thus, an inertial observer on one side of the wall will see
the domain wall as a sphere which accelerates towards the observer for
$T<0$, stops at $T=0$ at a radius ${k}^{-1}$, then accelerates away
for $T>0$.  We illustrate this construction in Fig.~\ref{vis}, where
we include the acceleration horizons to emphasize the causal
structure.  \vspace*{0.3cm}

Now, the repulsive effect of this vacuum domain wall is very similar
to the inflationary effect of a positive cosmological constant seen in
de Sitter space.  As we noted above, inflation provides an energy
source for the pair creation of black hole pairs in the early
universe.  Similarly, we would expect the repulsive gravitational
energy of the VIS domain wall to provide a mechanism for black hole
creation and indeed this was shown to be the case in \cite{cald}.  We
will now discuss this process of black hole pair creation in some
detail, because it will provide a prototypical example for our general
construction.

\subsection{Black Hole Pair Creation on a Domain Wall Background}

\begin{figure}[htb]
\hspace*{\fill} \vbox{\epsfxsize=11cm
\rotate[r]{\epsfbox{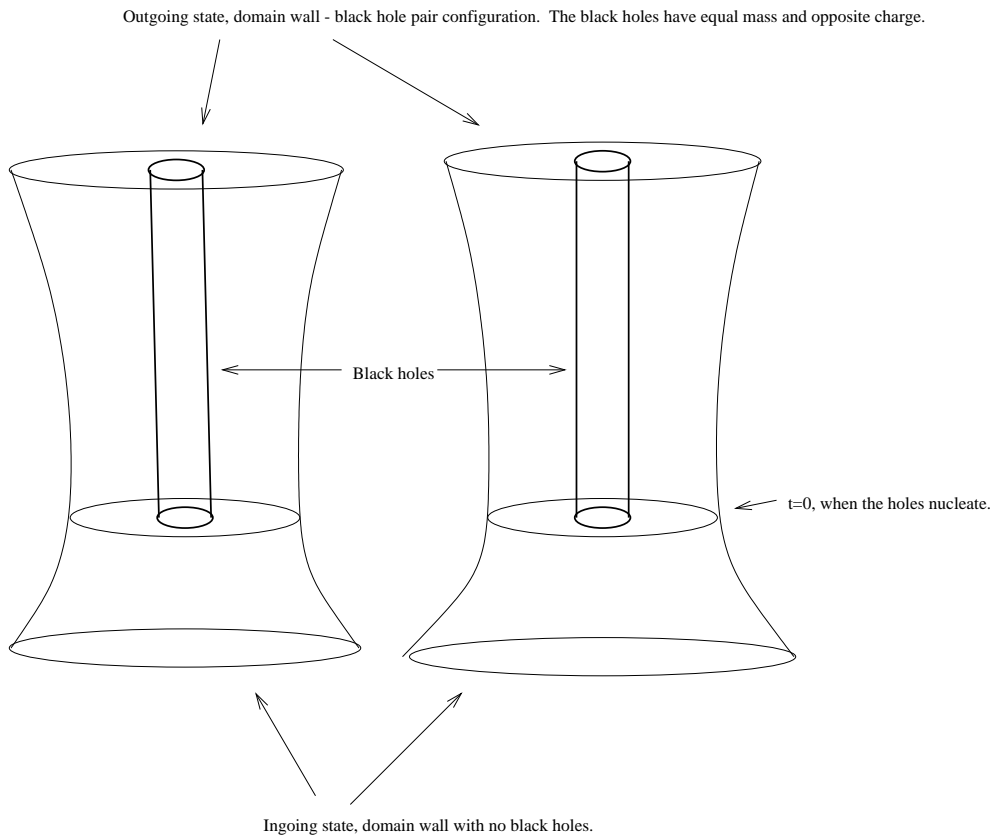}}}\hspace*{\fill}
\vspace*{0.3cm}
\caption[Pair creation of black holes by domain walls.]
{\small\sl A pair of black holes nucleated via the repulsive 
gravitational energy of the VIS domain wall.}
\label{bhpair}
\end{figure}

In \cite{cald} the creation rates of charged and uncharged black hole
pairs on a VIS background were calculated using the No-Boundary
Proposal.  Thus, amplitudes were calculated by first finding the
Euclidean action of the initial state (consisting of a single domain
wall with no black holes present), then finding the Euclidean action
of the final state (describing a domain wall with a black hole on each
side) and then applying equation (\ref{eq-pcr-qc}) to obtain the
correct rate.  The actual black hole creation process which was
studied is illustrated in Fig.~\ref{bhpair}. \vspace*{0.2cm}
 
\noindent (Actually, in \cite{cald} the 
only creation process studied was that of black holes which are
spherically centered on the domain wall, i.e., each black hole was
required to sit exactly in the center of each side of the original
spherical domain wall.  This is because the motion of a domain wall
which is spherically centered on a black hole is known in analytic
form.  Of course, one could also study the pair creation of black
holes which are `off center', but it is likely that numerical methods
would be required to simulate the exact wall motion after the black
holes were created.)

\begin{figure}[htb]
\hspace*{\fill} \vbox{\epsfxsize=11cm
\rotate[r]{\epsfbox{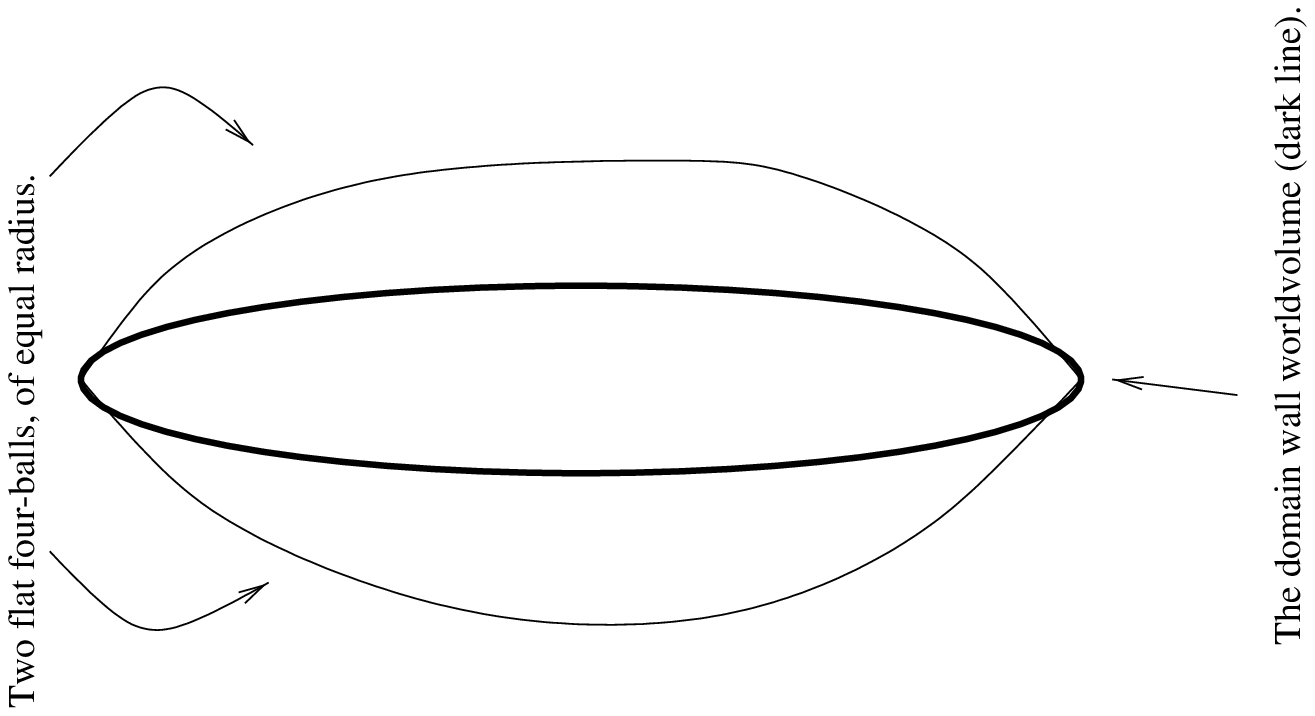}}}\hspace*{\fill}
\vspace*{0.3cm}
\caption[The lens instanton.]
{\small\sl The lens instanton is obtained by gluing two flat
  four-balls together along their respective $S^3$ boundaries.}
\label{lens}
\end{figure}

The instanton, or Euclidean section, of the VIS solution is very
similar to the $S^4$ instanton which mediates the creation from
nothing of a de Sitter universe.  This instanton allows us to
calculate the rate at which the initial state, with no black holes,
will be created from nothing.  Since the Lorentzian section of the VIS
configuration is just two portions of flat Minkowski spacetime glued
together, a natural `guess' for the Euclidean section is to take two
flat Euclidean four-balls and glue them together along a common
($S^3$) boundary.  When we do this we obtain the so-called `lens
instanton', which acquires its name from the fact that it looks rather
like a lens with a ridge of curvature running along the hemisphere
where the domain wall, $D$, is located, as illustrated in
Fig.~\ref{lens}. \vspace*{0.2cm}

Using equation (3.3) above, one calculates \cite{cald} that the total
Euclidean action of this instanton is
\begin{equation}
I_{D} = -\frac{\sigma}{2}{\int}_{\!\!D}d^{3}\!x\, \sqrt{h}
 = \frac{-1}{8{\pi}{\sigma}^2},
\end{equation}

\noindent where ${\int}_{\!D} d^{3}\!x \sqrt{h}$ 
is the volume of the domain wall worldsheet (the `ridge') on the
instanton, and we have used the fact that the radius, $r$, of each
four-ball is given in terms of the energy density as
\begin{equation}
r = \frac{1}{2{\pi}{\sigma}}.
\end{equation}

\begin{figure}[htb]
\hspace*{\fill} \vbox{\epsfxsize=8cm
\rotate[r]{\epsfbox{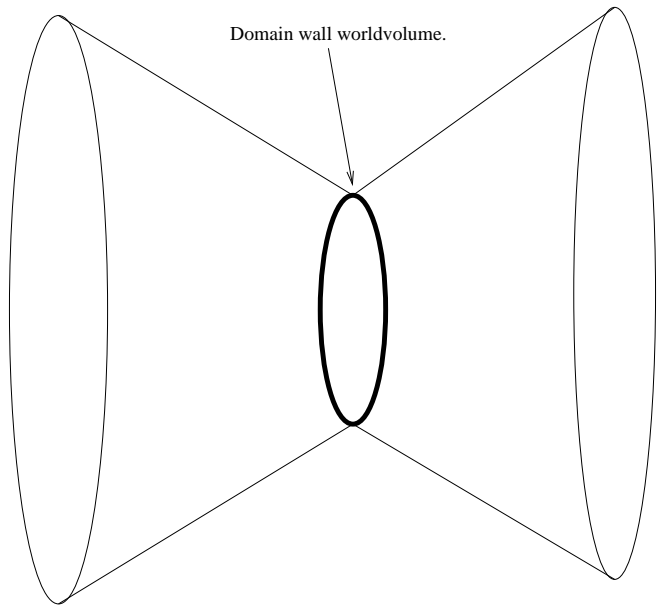}}}\hspace*{\fill}
\caption[A `yo-yo' instanton.]
{\small\sl Instanton (locally) for a 
negative energy density domain wall.}
\label{yoyo}
\end{figure}

\noindent Note that the energy density 
${\sigma} = 1/(2{\pi}r)$ is manifestly positive here because of the
sign of the extrinsic curvature as one moves across the domain wall
worldsheet.  More explicitly, the extrinsic curvature of a sphere of
radius $r$ in flat space is of course ${\pm}1/r$, where the sign is
determined by whether one is calculating relative to the outward or
inward normal to the surface.  As one approaches the VIS domain wall
from one side, the 3-spheres are locally `expanding' and so the
extrinsic curvature is given as $K_{ij}^{+} = +(1/r)h_{ij}$.
Likewise, as one recedes from the domain wall on the other side the
3-spheres are locally contracting, and so the extrinsic curvature has
the sign $K_{ij}^{-} = -(1/r)h_{ij}$.  Using condition (2) of the
Israel matching conditions described above we thus see that the energy
density satisfies $(1/r)h_{ij} - (-(1/r)h_{ij}) =
4{\pi}{\sigma}h_{ij}$, from which Eq.~(3.10) follows.  The reason we
are emphasizing this point is that one can reverse the sign of the
energy density simply by considering domain wall worldsheets where the
extrinsic curvature has the opposite behavior as one moves through
the domain wall.  For a domain wall of negative energy density, as one
approaches the wall from one side the spheres will be locally {\it
  contracting}, and when one leaves the wall from the other side the
spheres will start expanding again.  Thus, locally the Euclidean
section of such a domain wall would look rather like a `yo-yo', with
the domain wall worldsheet running along the groove for the string, as
illustrated in Fig.~\ref{yoyo}.  \vspace*{0.2cm}

We now need to describe the Euclidean section of the `final state',
which consists of a pair of black holes moving relative to the domain
wall.  As discussed in \cite{cald}, the motion of black holes relative
to a thin domain wall was worked out long ago by Berezin et
al.~\cite{ber} and Hiscock \cite{cock}, for the case where the black
holes are spherically centered in the middle of each side of the
domain wall.  The exact equations were presented in \cite{cald} and
are not important for our analysis here.  Here we simply sketch the
main physical properties of a black hole - domain wall configuration.

If the created black holes each carry a single $U(1)$ charge, then the
only physical parameters in the problem are the masses of the holes
(assumed to be equal), the charges of the holes (assumed to be
opposite and equal) and the energy density of the wall.  Since the
domain wall is repulsive and the black holes attract each other, there
are basically three cases:

\noindent {\bf Case 1}: The repulsive energy density of the 
wall overwhelms the attractive force between the holes and the black
holes continue to move apart after they have been created.
 
\noindent {\bf Case 2}: The repulsive energy of the wall exactly 
counterbalances the attraction between the black holes and the final
configuration is in static equilibrium.

\noindent {\bf Case 3}: The attractive force
between the holes is greater than the repulsive force of the domain
wall and the holes eventually crash together.

While all of these possibilities are in principle allowed, we will
focus on Case 2 since in that situation the construction of the
instanton is much simpler.  It will be clear, however, that everything
we say here will also go through for the other cases.  As discussed in
\cite{cald}, a solution always exists for the motion of a static
domain wall relative to a black hole.  If the black hole has mass $m$
and charge ${\pm q}$, then the (totally umbilic) domain wall
hypersurface lies at the constant radius $r_{\rm s}$ given by
\[
r_{\rm s} = {3 \over 2} m \Big[ 1 + 
\sqrt{1 - {8 \over 9} {q^2 \over m^2}} \Big].
\]

\noindent Thus the final state, 
consisting of two black holes of opposite charge separated by a static
spherical domain wall, is obtained by taking two copies of
Reissner-Nordstr{\"o}m, cutting each copy along a timelike cylinder at
$r = r_{\rm s}$, then gluing the two solid interiors of the cylinders
along the domain wall hypersurface.  The Euclidean section for this
configuration is therefore obtained by taking two `cigar instantons'
(for Reissner-Nordstr{\"o}m spacetime), snipping each cigar along the
hypersurface $r = r_{\rm s}$ (taking care to keep the `tip' of each
cigar, where the black hole horizons are), then gluing the two ends of
the cigars together along this surface.  The action of this instanton
is calculated to be
\begin{equation}
I_{\rm Dbh} = -2 \pi \sigma r^2 \beta_{\rm RN}
 {\widetilde f}^{1/2} |_{r_{\rm s}}
+ q^2 \beta_{\rm RN} \Big( {1 \over r_{+}} - {1 \over r_{\rm s}} \Big),
\end{equation}
where ${\beta}_{\rm RN}$ is the period of the Reissner-Nordstr{\"o}m
cigar instanton, $r_{+}$ denotes the outer black hole horizon radius,
and $\widetilde f = 1 - 2m/r + q^2/r^2$.

Given this, we can apply the No-Boundary Proposal and
Eq.~(\ref{eq-pcr-qc} to obtain the probability that static black hole
pairs (of mass $m$ and charge ${\pm q}$) will be nucleated by a VIS
domain wall:
\begin{equation}
P = \frac{P_{\rm BH}}{P_{\rm noBH}} = 
\exp\Big[ -{1 \over 8 \pi \sigma^2} 
+ 2 \pi \sigma r_{\rm s}^2 \beta_{RN} \widetilde f^{1/2}
- q^2 \beta_{RN} \Big( {1 \over r_{+}} - {1 \over r_{\rm s}} \Big)
\Big] 
\end{equation}

Similar probabilities are obtained when the created holes are allowed
to accelerate relative to the domain wall (the only subtlety in the
non-static situation is that there are non-trivial matching conditions
which the Euclidean sections of the black hole - domain wall
configurations must satisfy in order to be well-defined instantons).
Clearly, the probability is heavily suppressed when the wall energy
density $\sigma$ is small, as would be expected.

Using the No-Boundary Proposal, we have described the calculation of a
(generic) probability that stationary black hole pairs will be created
in the presence of a VIS domain wall.  However, this approach has not
told us how to construct an imaginary time path connecting the initial
and final states.  We will now show how to construct such a path which
contains an off-shell wormhole fluctuation of arbitrarily large energy
density, but arbitrarily small action.

\section{How to Build Interpolating Paths}

\label{sec-interpol}

When one uses the No-Boundary Proposal to calculate a tunneling
amplitude in a cosmological scenario, what one does conceptually is
calculate the rate at which one universe will annihilate, and another
universe will be created, in its place.  Here, we will describe how to
`patch up' this picture by performing surgery on the no-boundary
instantons to obtain a connected path which connects the ingoing and
outgoing universes.

The surgery is actually quite simple, and it is closely related to the
operation of taking the connected sum of two manifolds in topology.
The only subtlety here is that we want to take the connected sum in
such a way that the surface along which the manifolds are joined
satisfies the Israel matching conditions, so that we can interpret the
matching surface as a virtual domain wall.  The resulting manifold
will then be an off-shell history which connects the ingoing state to
the outgoing state.

\begin{figure}[htb]
\hspace*{\fill} \vbox{\epsfxsize=11cm
\rotate[r]{\epsfbox{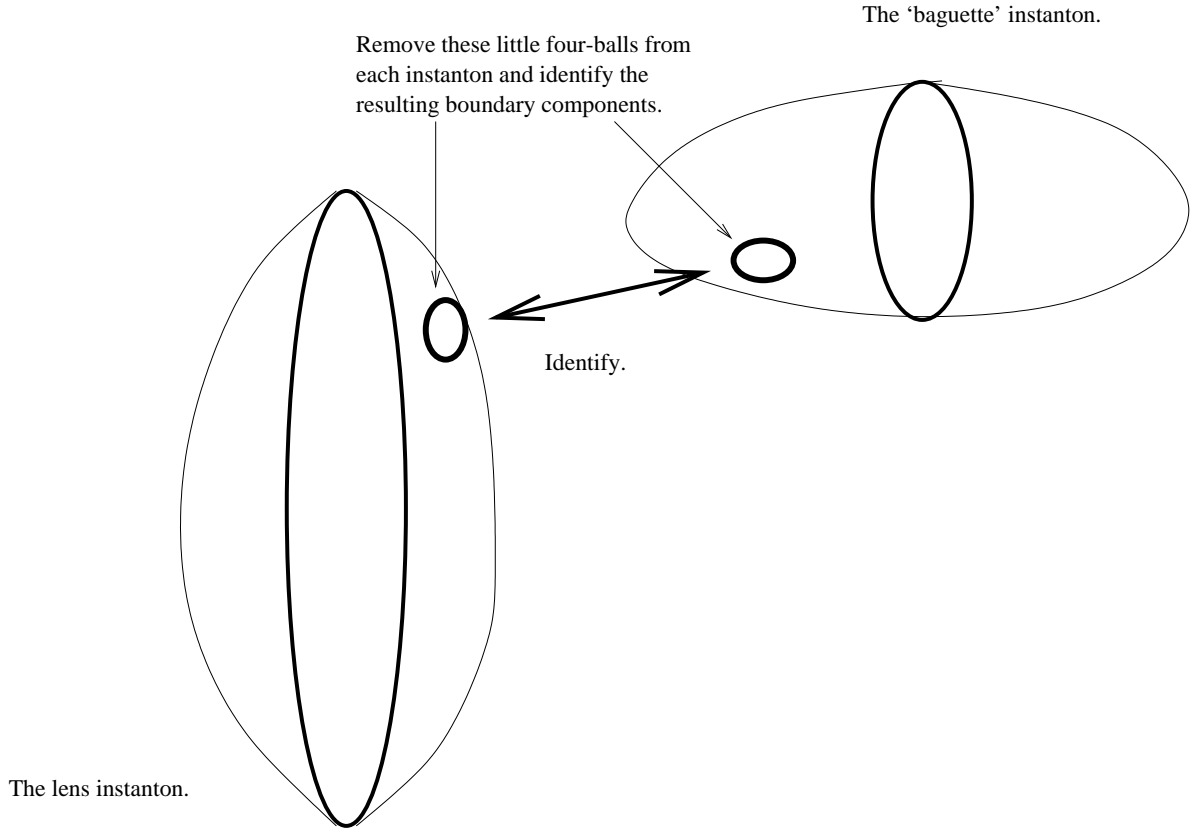}}}\hspace*{\fill}
\caption[The patching procedure.]
{\small\sl Gluing the instantons together along the virtual domain
  wall, or thin wormhole.}
\label{vis2}
\end{figure}

In order to have an explicit example where we can implement this
construction, let us return to the above scenario where black holes
are created on the VIS background.  As we saw above, the instanton for
the initial state was half of the `lens instanton' ($M_L$), which was
obtained by gluing two flat four-balls together.  Likewise, the
instanton for the final state was half the `baguette' instanton
($M_B$), obtained by gluing the ends of two cigar instantons together.
We now patch these two instantons together by first removing a (solid)
four-dimensional ball, $B^{4}(\delta)$, of radius $\delta$, from each
of the instantons.  The boundary components left behind once we remove
these small balls are then totally umbilic three-spheres of equal
radius; we now join the two instantons together along these
three-spheres in such a way that the joining surface satisfies the
Israel matching conditions.  This construction is illustrated in
Fig.~\ref{vis2}. \vspace*{0.1cm}
In this way we obtain an interpolating history which models the decay
of of VIS domain wall to a VIS - black hole system.  We will show
below that the action for this path will be very similar to the action
difference of the two instantons used in the No-Boundary Proposal.

However, before turning to this calculation we should point out that
the energy density of the virtual wormhole must be {\em negative}.
This follows easily from the analysis presented above: As one approaches
the wall hypersurface from one side, neighboring umbilic three-spheres
are shrinking, and as one leaves the wall on the other side the
three-spheres are expanding.  Thus, the three-sphere corresponding to
the domain wall worldsheet (which we interpret as a `virtual' domain
wall of topology $S^2$ which appears from nothing, expands briefly,
then annihilates) has negative energy density.  Indeed, the energy
density, which we denote by ${\bar \sigma}$, is set by the scale of
the virtual domain wall; it is calculated to be
\begin{equation}
{\bar \sigma} = \frac{-1}{2{\pi}{\delta}}.
\label{eq-sigma-delta}
\end{equation}

Regions of negative energy are perfectly allowed for off-shell paths
such as the ones we are constructing here. There are two reasons why
one should not attempt a similar prescription using real domain walls
(on-shell paths): First, it would require assumptions about the matter
fields -- they would have to allow domain walls.  This would make the
analysis less general. Second, there is no sensible classical field
theory which can give rise to domain walls of negative energy density,
because this would destroy vacuum stability.  Of course, this may seem
confusing given that wormholes supported with large amounts of
negative energy density are considered every day by the `wormhole
engineers' \cite{visser}, who need the negative energy density in
order to make the wormholes traversable.  However, these engineers use
quantum mechanical processes (such as the Casimir effect) to construct
regions with large negative energy density.  We will not consider such
issues here.

Given this off-shell resonance, we now want to calculate the action
and get a tunneling amplitude.  This calculation is fairly
straightforward once we notice several elementary points.

First of all, there are no volume contributions to the action in this
particular scenario, and so we don't have to worry about the fact that
we have removed four-balls from the original instantons (as expected,
the only contribution there will come from the virtual wormhole
itself).  Actually, as we shall show in a moment that one never has to
worry about the volume contributions (even when there is a
cosmological constant, for example) since the volume terms from the
removed balls will always cancel each other.

Second, the virtual domain wall hypersurface is crucial because it
ensures that the initial and final instantons will have {\it opposite}
relative orientations.  This means that the two actions will appear
with the correct relative signs.  

In order to understand this orientation change, it is useful to
consider what happens when you try to extend a ($C^{0}$) tetrad field
through the surface of the domain wall.  To be concrete, assume that
the tetrad fields on each of the little balls removed from the the
instantons are comprised of the vectors naturally associated with
polar normal coordinates on each ball.  That is to say, three vectors
of a tetrad are taken to be (angular) coordinates tangent to the
boundary of a ball, and the fourth vector is a (radial) vector normal
to the boundary of a ball.  Now, if we want the tetrads on each ball
to `match' at the surface of the domain wall, it follows that we {\it
  must} take the normal component of the tetrad to be inward pointing
on one ball and outward pointing on the other.  Thus, as we move from
one instanton to the next we reflect about one leg of the tetrad,
i.e., we reverse the orientation.

Indeed, this is precisely why we don't have to worry about the removed
volume contributions.  If the removed four-balls have equal, but
opposite action, then the final difference (after they have been
removed) is equal to the original No-Boundary Proposal difference.  Of
course, this cancellation actually only works when the four-balls have
{\it exactly} equal action contributions.  One might imagine a
situation where, for example, the cosmological constant `jumped' to a
different value during the decay process (as you moved from the
initial instanton to the final instanton across the virtual domain
wall worldsheet).  In such a scenario, the actions of the two removed
four-balls would not be equal and so there would be an extra
correction term to the no-boundary result.  However, we will not
consider such complications in this paper.

Given these comments, it is now clear that the final action, $I_T$,
for our interpolating path is just the original no-boundary difference
plus a small correction term coming from the virtual domain wall:
\begin{equation}
I_{T} = \frac{1}{2} I_{\rm Dbh} - \frac{1}{2} I_{D}
 + \frac{1}{8{\pi}{\bar \sigma}^2}
\label{eq-halfaction}
\end{equation}
where $I_{\rm Dbh}$ and $I_D$ are given by (3.11) and (3.9)
respectively and the correction term involving ${\bar \sigma}$ appears
with a plus sign because the virtual domain wall has a negative energy
density.

Taking the point of view of the bounce approach (see
Sec.~\ref{ssec-instapproach}) one should complete the newly connected
half-instantons with their mirror image to obtain a bounce, in which
one starts with the initial spacelike section, goes though a final one
and back to the initial type at the other end of the Euclidean
geometry. This requires another surgery in Fig.~7, with a second
virtual domain wall allowing the transition from the baguette back to
the lens. The total action will obviously be twice that in
Eq.~(\ref{eq-halfaction}). Using Eq.~(\ref{eq-sigma-delta}), it may be
written in the form
\begin{equation}
I(\delta) = I_0 + \frac{1}{2} \rho \delta^2,
\end{equation}
where 
\begin{equation}
I_0 = I_{\rm Dbh} - I_{\rm D},~~~~
\rho = 2\pi.
\label{eq-gauss}
\end{equation}
The transition rate will be given by
\begin{eqnarray}
\Gamma & = & \int_{-\infty}^{\infty} d\delta\,  e^{-I} \\
 & = &  e^{-I_0} \int_{-\infty}^{\infty} d\delta\, 
 e^{-\frac{1}{2} \rho \delta^2 } 
\label{eq-pathint2} \\
& = & e^{-I_0}.
\label{eq-prefexp2}
\end{eqnarray}

We demand that the disconnected geometry be excluded from this path,
and we assume that the connecting virtual domain wall must have a
diameter of at least the Planck length. This will restrict the range
of integration in Eq.~(\ref{eq-pathint2}) to the regions of more than
one standard variation, and will therefore reduce the value of the
prefactor from 1 to about $1/3$.  Given the exponential suppression of
black hole pair creation, this change is negligible. Therefore, the
requirement for connected interpolating geometries will not alter the
no-boundary approach to cosmological pair creation significantly: The
exponent will be unchanged, and the prefactor, which is usually
neglected anyway, will still be of the same order of magnitude. Our
approximation breaks down only for Planck-scale background geometries,
when the semi-classical approach fails in any case.

Of course, one can also apply this construction to other gravitational
tunneling phenomena, such as black hole pair creation on a
cosmological background.  There again, as long as the cosmological
constant is conserved in the decay process, the final answer will be
the original no-boundary result with a reduced prefactor.

One might object to the use of off-shell paths. However, they are not
only an essential part of the formal path integral formalism, but they
give a significant contribution to the saddlepoint approximation --
after all, the actual saddlepoint solution forms a set of measure zero
in the saddle point contour. Crucially, we demonstrated that the
interpolating off-shell paths we considered are in fact arbitrarily
small perturbations of the saddlepoint solution. Thus, if the
saddlepoint is excluded from the integral, these geometries will still
dominate.  Therefore we stress that on the contrary, our proposal
offers a consistent way of including the effects of spacetime foam in
semi-classical calculations.

We should also point out here that our construction is in many
respects similar to the earlier work of Farhi, Guth and Guven
\cite{fgg}, who were studying the rate at which a new universe could
be created in the laboratory.  In their approach, one constructs an
interpolating instanton by gluing the two instantons together to
obtain a `two-sheeted' pseudomanifold.  The basic idea there is that
the ingoing state is on one sheet, and the outgoing state is on the
other sheet.  This approach was recently employed by Kolitch and
Eardley, who studied the decay of vacuum domain walls.  Interestingly,
they found that the rate calculated using the Farhi, Guth, Guven (FGG)
technique is identical to the rate calculated using the No-Boundary
Proposal. Thus, it would seem that attempts to construct some
interpolating geometry (in situations where no connected, on-shell
path exists) will always lead back to the no-boundary ansatz.

The current debate about the correct boundary conditions on the wave
function of the universe~\cite{argue} centers on the question of
whether the Hartle-Hawking No-Boundary Proposal, or the Tunneling
Proposal favored by Linde~\cite{Lin84b}, Vilenkin~\cite{Vil86} and
others, should be used to describe the creation of a universe from
{\it nothing}. We emphasize here that the use of the No-Boundary
Proposal for the tunneling processes on an {\em existing} background
is not called into doubt by either Hawking or Linde. Here we have
aimed to elucidate the reasons for the success of the No-Boundary
Proposal for such processes.

Finally, in light of recent work \cite{alex} concerning the possible
role of singular instantons in describing tunneling phenomena, we would
like to emphasize that the interpolating paths which we have constructed
here are in no way singular.  Rather, these trajectories simply contain
off-shell fluctuations which may be regarded as distributional sources
of negative energy density.

\section{Summary}

In this paper we have aimed to justify the use of compact instantons
for the semiclassical description of tunneling processes on
cosmological backgrounds. We found connected off-shell interpolating
geometries which can be viewed as small perturbations of the
disconnected on-shell instantons. Therefore they can dominate the path
integral. Since the disjoint solutions are linked by a virtual domain
wall in our approach, the total action of the interpolating geometry
will be the {\em difference} between the instanton actions, and we
recover the result obtained from the No-Boundary approach.

Over the years, black hole pair creation has been investigated on a
variety of backgrounds. Usually, one finds a Euclidean bounce solution
which includes spacelike sections with and without black holes. The
difference between the bounce action and the action of the Euclidean
background solution is calculated. The pair creation rate is obtained
as the exponential of minus this action difference, as in
Eq.~(\ref{eq-pcr-usual}).

This method fails for the cosmological and domain wall backgrounds we
have considered, because there is no single Euclidean solution that
will interpolate between the initial and final spacelike sections.
Instead, one can construct two nucleation geometries (half-bounces,
which are continued in a Lorentzian direction rather than back to the
initial spacelike section), which both describe the nucleation of a
universe from nothing: one with black holes, the other without.
Therefore, the pair creation problem can be attacked within the
framework of quantum cosmology. This leads to a prescription for the
pair creation rate of cosmological black holes.

This `quantum cosmological' approach rests on the assumption that each
Hubble volume in an inflationary universe can be regarded as having
been nucleated independently. In quantum cosmology, one constructs a
wave function of the universe. The square of its amplitude gives a
probability measure. By taking the ratio of the probability measures
assigned to the two instanton types, one can calculate the relative
probability for a Hubble volume to nucleate with a black hole pair,
compared to an empty Hubble volume. Unless the cosmological constant
is of Planck order, this number is small, and can be interpreted as a
pair creation rate in the natural length and time scale, the Hubble
scale.

One then uses the Hartle-Hawking No-Boundary Proposal to determine the
wave function semi-classically. According to this proposal, the
probability measure will be given by the exponential of minus twice
the real part of the instanton action, which is equivalent to the full
bounce action of the usual pair creation treatment. The ratio of these
two exponentials is, of course, equivalent to a single exponential of
the action difference.  Thus, one recovers the usual prescription for
the pair creation rate, as far as is possible given the fundamental
differences between the cosmological and the non-cosmological
situations. This is an important test for consistency, especially
since the instanton actions reflect the geometric entropy of the
nucleated spacetimes.  Schwarzschild-de~Sitter space has a lower
geometric entropy than de~Sitter space. Thus the instanton actions
reflect the physical necessity that transitions in the direction of
lower entropy are suppressed.

While this application of the No-Boundary Proposal to decay processes
in quantum cosmology would seem to be intuitively justified by these
arguments, it seemed to rely on two disconnected instantons, instead
of describing the transition though a single Euclidean geometry
connecting the initial state to the final state.

In this paper, we showed that the exact saddlepoint solution is a
special disconnected geometry in a generic class of connected ones, in
which the instantons are patched together using virtual domain walls.
These off-shell geometries can be arbitrarily close to the saddlepoint
and contribute to its domination of the path integral. The exclusion
of disconnected geometries therefore has no fundamental effect on the
formalism. The exponential suppression of the pair creation process is
left unchanged, and we estimated that the prefactor would be
diminished by a factor of a third.

One could also use our method to study other decay processes which are
expected to occur in the early universe.  A natural candidate process
would be the decay of vacuum domain walls by quantum tunneling
recently discussed by Kolitch and Eardley (\cite{shawn1},
\cite{shawn2}).  This process is important because it provides another
decay mode capable of eliminating the unwanted gravitational effects
of domain walls in the early universe.  It is also possible to generalize the
Kolitch-Eardley analysis to supergravity domain walls (such as the
D8-branes of the IIA theory) which arise in the low-energy limit of
string theory \cite{d8}.  Research on these and related problems is
currently underway.

\mbox{}\\
{\noindent \bf Acknowledgments}\\

The authors would like to thank Doug Eardley, Gary Gibbons, Stephen
Hawking, Ted Jacobson, Shawn Kolitch, Andrei Linde, Robert Mann and
Don Marolf for useful conversations.  R.B.\ was supported by
NATO/DAAD.  A.C.\ was supported by Pembroke College, University of
Cambridge.


\end{document}